\documentclass[prd,floatfix, nofootinbib, preprint]{revtex4-1}
\usepackage{amssymb}
\usepackage{slashed}
\usepackage{amsmath}
\usepackage{braket}
\usepackage{float}
\usepackage[utf8]{inputenc}
\usepackage{graphicx}
\usepackage{subfigure}
\usepackage{latexsym}
\usepackage{epic}
\usepackage{eepic}
\usepackage{epsfig}
\usepackage{color}
\usepackage{natbib}

\begin{document}
\title{Probing Leptoquark and Heavy Neutrino at LHeC}
\author{Sanjoy Mandal}
\email{smandal@imsc.res.in}
\affiliation{The Institute of Mathematical Sciences,
C.I.T Campus, Taramani, Chennai 600 113, India}
\affiliation{Homi Bhabha National Institute, BARC Training School Complex,
Anushakti Nagar, Mumbai 400085, India}
\author{Manimala Mitra}
\email{manimala@iopb.res.in}
\affiliation{Institute of Physics, Sachivalaya Marg, Bhubaneswar 751005, India}
\affiliation{Homi Bhabha National Institute, BARC Training School Complex,
Anushakti Nagar, Mumbai 400085, India}
\author{Nita Sinha}
\email{nita@imsc.res.in}
\affiliation{The Institute of Mathematical Sciences,
C.I.T Campus, Taramani, Chennai 600 113, India}
\affiliation{Homi Bhabha National Institute, BARC Training School Complex,
Anushakti Nagar, Mumbai 400085, India}
\preprint{IP/BBSR/2018-9}

\begin{abstract}

We explore leptoquark production and decay for the $\tilde{R}_{2}$ class of models at the proposed $e^{-} p$ collider LHeC, planned to operate with 150 GeV electron and 7 TeV proton beams.  In addition to the  coupling of 
leptoquark with the lepton and jet, the model also has right handed neutrinos, coupled to the leptoquark. We analyse the collider signatures of a number of final states, that can originate from leptoquark decay into the standard model particles, as well as, 
the final states that originate from further decay of the heavy
neutrinos, produced from leptoquark. We find that the final state $\ell^{-}+\text{n-jets}\,(1\leq 
\text{n} \leq 2)$
has the largest discovery prospect, more than $5\sigma$ with only few fb of 
data to  probe a leptoquark  of mass 1.1 TeV, even with a generic set of cuts. The significance falls sharply 
with increasing 
leptoquark mass. However, with 100\,$\rm{fb}^{-1}$ of data,  a  5$\sigma$ discovery for  leptoquarks 
of mass upto 1.4 TeV is still achieveable. Also for the same luminosity, final state $\bar{b}\ell^{+}\tau^{-}+\text{n-jets}\,(\text{n}\geq 2)+\slashed{E}_{T}$, resulting from the cascade decay of the leptoquark to an $\bar{t}$ and right handed neutrino, followed by further decays of $\bar{t}$ and the neutrino, is expected to yield a rather large number of events~($\approx 180$).
\end{abstract}

\maketitle
\section{Introduction}
Discovery of physics beyond the Standard Model (SM) continues to be the aim of 
most high energy physicists today. Inspite of the so far unsuccessful
direct searches for new particles or clear indications for existence of any 
kind of new physics at the Large Hadron Collider (LHC), issues of existence of dark 
matter, baryon asymmetry of 
the universe, 
non-zero neutrino masses etc., are 
compelling enough for one to believe that there must exist physics beyond the 
SM.  A number of beyond Standard Model extensions have been proposed in the 
literature. Minimal supersymmetric standard model, large extra dimensions, 
seesaw models of 
neutrino mass generation, leptoquarks and grand unified theories, are a few 
among them. 
 
Leptoquarks (LQs) are hypothetical particles, which make leptons couple 
directly to quarks and vice versa~\cite{PatiSalamFourthColor,Buchmuller:1986zs,Buchmuller:1986iq}. In the Pati-Salam model, they emerged from 
the unification of quarks and leptons~\cite{PatiSalamMatterUnification}. They also exist in grand unification 
theories based on SU(5)~\cite{GeorgiGlashowSU5} and SO(10)~\cite{Georgi:1974my,Fritzsch:1974nn,Mohapatra:1979nn,Wilczek:1981iz,Melfo:2010gf,Babu:1992ia,Bajc:2002iw,Joshipura:2011nn,Dev:2009aw,Bajc:2004fj}. They are also expected to exist at TeV 
scale in extended technicolour models~\cite{DimopoulosSusskindTechnicolor,DimopoulosTechnicolor,FarhiSusskindTechnicolor,GeorgiGlashowTechnicolor}. LQs can be either of scalar or 
vector nature. Using the SM representation of quarks and leptons, all possible 
LQ states can be classified, with six scalar and six vector LQ multiplets under 
the SM gauge group \cite{LQReview}. 
Among the different classes, the scalar LQ $\tilde{R}_2$ is   
interesting, as it is one of the multiplets that allows for matter stability~\cite{Arnold:2012sd}. 
Moreover, it also couples to right handed neutrinos (RH neutrinos).
The RH neutrinos can generate  light 
neutrino masses through   seesaw mechanism \cite{MinkowskiSeeSaw,MohapatraSeeSaw,GellmannSeeSaw,YanagidaSeeSaw,
SchechterValleSeeSaw,SchechterValle2SeeSaw,WeinbergSeeSaw,Weinberg2SeeSaw,MaggSeeSaw,ChengSeeSaw,FootSeeSaw,inverseseesaw1,inverseseesaw2,inverseseesaw3}.
In seesaw, light neutrino masses are generated through $d=5$ lepton number violating operator~\cite{WeinbergSeeSaw}. The high scale UV completed models  include gauge singlet RH neutrino (type-I, inverse seesaw), $SU(2)_L$ triplet scalar and 
fermion (type-II, and type-III). 

In the case of type-I seesaw \cite{MinkowskiSeeSaw,MohapatraSeeSaw,GellmannSeeSaw,YanagidaSeeSaw}, the light neutrinos acquire masses through mixings 
with additional Majorana RH neutrinos. To account for the tiny 
neutrino masses, the mass scale of these Majorana neutrinos has to be very 
close to the gauge coupling unification scale, in which case these massive 
RH  neutrinos will remain inaccessible at LHC as well as, at other 
near future colliders.  For the present and near future colliders to be able to probe the RH neutrinos, their masses have to be within the experimental reach, and the mixing with the active-neutrinos (referred as active-sterile mixing) has to be sizable. TeV scale RH neutrinos with substantially large active-sterile mixings are however possible to accommodate in type-I seesaw if cancellation exists in the light neutrino mass matrix \cite{Mitra:2011qr}.  
The inverse 
seesaw mechanism \cite{inverseseesaw1,inverseseesaw2,inverseseesaw3}  is another  such scenario, where TeV scale or even smaller RH neutrino masses with sizeable active-sterile mixing can exist. In this scheme,
in addition to the SM
particles there are gauge  singlet neutrinos, with opposite lepton numbers 
(+1 and 
-1). The light neutrino mass matrix is given in terms of the Dirac neutrino mass term, 
$m_D\sim Y_\nu v$ (with $v$ being the electroweak vev and $Y_\nu$, a generic 
Yukawa coupling), the heavy  neutrino mass scale $M_R$ and, a small 
lepton number violating ($\Delta L=2$) mass term $\mu$, which ensures that $M_R$ 
scale remains close to TeV or less, with order one Yukawa coupling. The light neutrino mass matrix in this case is: $m_\nu\sim 
(m_D^2/M_R^2)\mu$. While the heavy neutrino states may lie within the kinematic reach of LHC, 
their production cross section falls rapidly with increasing  
masses and smaller active-sterile neutrino mixing. Large active-sterile mixing is possible to obtain in other seesaw secenarios as well, such as, linear seesaw~\cite{Linearseesaw1,Linearseesaw2}, extended seesaw~\cite{Extendedseesaw1,Extendedseesaw3,Kang:2006sn,Majee:2008mn,Parida:2010wq}. A substantial 
rise in the production cross section of the RH neutrinos is feasible 
in the presence of LQs. This has been explored recently for LHC in \cite{Das:2017kkm} for inverse seesaw, where a number of final states have been analysed in detail.  Leptoquark models have also been tested recently for fitting the IceCube events~\cite{Dey:2017ede}. 
For the heavy neutrino searches at LHeC in inverse seesaw model, see \cite{Subhadeep} and for the LNV signal at LHeC, see~\cite{Mondal:2015zba}. Similar studies for heavy neutrino searches also been carried out in~\cite{Lindner:2016lxq,Antusch:2016ejd}.

In this work, we consider a particular type of scalar leptoquark $\tilde{R}_{2}$ 
which transforms as $\tilde{R}_{2}\in (3,2,\frac{1}{6})$ under the SM gauge 
group $SU(3)_{c}\times SU(2)_{L}\times U(1)_{Y}$ and for the RH neutrino, we adopt model independent framework.
$\tilde{R}_{2}$
is a genuine LQ with fermion number $F = 3B+L = 0$. 
The colour charge of the LQ will enable in their copious production at 
LHC. Moreover, at $e^{-}p$ colliders like LHeC, they can be resonantly produced. The 
LHeC is a proposed $e^{-}p$ collider in the TeV regime after HERA, supposed to be 
built in the LHC tunnel \cite{LHeCDesignReport}. LHeC will use a newly built 
electron beam of 60 GeV, up to possibly 150 GeV, to collide with the intense 7 
TeV proton beam of the LHC. LHeC is expected to operate with 100 $\rm{fb}^{-1}$ 
integrated luminosity, and 
is complementary to the $pp$ collider LHC \cite{ComplementLHC}. The RH neutrino, being coupled to the LQ, can be produced from LQ decay. 
The decay of LQ into a lepton and a jet, and the decay of RH neutrino in different SM states give rise to a plethora of model signatures, that we study in detail. We show that among all the final states, 
$\ell^{-}+\text{n-jets} (1 \leq \text{n} \leq 2)$ has the highest LQ discovery prospect, even with generic sets  of cuts.  Additionally, we also carry out an in-depth analysis for few other channels that arise due to the decay of a heavy neutrino. We show that with 
judiciously applying selection cuts the channels $\ell^{-}+\text{n-jets} ( \text{n} \geq 3)$, and $\ell^{+}\tau^-\bar{b}+\slashed{E}_T+\text{n-jets} (\text{ n} \geq 2) $ can be made background free. 

The discussion of the paper goes as follows: in Section.~\ref{Model}, and Section.~\ref{Constraints on Leptoquark Couplings},  we review the model and the theory constraints. Following that, in Section.~\ref{Leptoquark Production}, we discuss the production and decay of 
LQ at LHeC. In the subsequent sections, Section.~\ref{Collider Analysis}, Section.~\ref{Signals and Background} and Section.~\ref{Signal Strength for Higher LQ Mass}, we present a detailed collider analysis and discuss the discovery prospects of different final states. 
Finally, in Section.~\ref{Conclusions}, we summarize. 

\section{Model}
\label{Model}
We consider the scalar LQ $\tilde{R}_2$ charged as (3,2,1/6) 
under SM gauge group.  In the presence of the RH neutrinos $N_R$, the 
LQ has additional interaction \cite{LQReview,Buchmuller:1986zs,Buchmuller:1986iq},
\begin{align}
 \mathcal{L}=-Y_{ij}\bar{d}_{R}^{i}\tilde{R}_{2}^{a}\epsilon^{ab}L_{L}^{j,b}+
 Z_{ij}\bar{Q}_{L}^{i,a}\tilde{R}_{2}^{a}N_{R}^{j}+h.c., 
 \label{eq1}
\end{align}
where $i,j=1,2,3$ are flavor indices and $a,b=1,2$ are $SU(2)_L$ indices.
We assume that there are three right-chiral neutrinos $N_{R}^{j}\, (j=1,2,3)$, 
$Y_{ij}$ and $Z_{ij}$ are the elements of arbitrary complex $3\times 3$ Yukawa 
coupling matrices. Note that,  $\tilde{R}_{2}$  comprises two LQs. One 
has $Q=\frac{2}{3}$, and the other has $Q=-\frac{1}{3}$. Upon expansion, the Lagrangian becomes
\begin{align}
 &\mathcal{L}=-Y_{ij}\bar{d}_{R}^{i}e_{L}^{j}\tilde{R}_{2}^{2/3}+(YU_{\text{PMNS}})_{ij}\bar{d}_{R}^{i}\nu_{L}^{j}
 \tilde{R}_{2}^{-1/3}+\\ \nonumber
 & (V_{\text{CKM}}Z)_{ij}\bar{u}_{L}^{i}N_{R}^{j}\tilde{R}_{2}^{2/3}+Z_{ij}\bar{d}_{L}^{i}
 N_{R}^{j}\tilde{R}_{2}^{-1/3}+h.c.,
 \label{eq2}
\end{align}
where the superscript of LQ fields denotes electric charge of a given $SU(2)_L$ doublet component of $\tilde{R}_{2}$,
$U_{\text{PMNS}}$ and $V_{\text{CKM}}$ are Pontecorvo-Maki-Nakagawa-Sakata (PMNS) and Cabibbo-
Kobayashi-Maskawa (CKM) matrices. At the  $e^- p$ machine  
$\tilde{R}_{2}^{\frac{1}{3}}$ can not be resonantly produced. Hence, the expected cross-section  for  the production of $\tilde{R}_{2}^{\frac{1}{3}}$
is  small. Therefore, in this work, we consider only  
$\tilde{R}_{2}^{\frac{2}{3}}$ and study its production.

The charged current and neutral current interactions of the 
RH neutrinos are parametrized in a model independent way as follows, 
\begin{align}
-\mathcal{L}_{CC} = \frac{g}{\sqrt{2}} W^{-}_{\mu}\bar{\ell} \gamma^{\mu} P_{L} V_{\ell j} N_j+{\rm H.c.},
\end{align}
and
\begin{align}
-\mathcal{L}_{NC}  = \frac{g}{2 \cos\theta_w}  Z_{\mu} \left\{  (U^\dag_{PMNS}
V)_{ij}{\bar{\nu}}_i \gamma^{\mu} P_L N_j
  + {\rm H.c.} \right\}
\label{NC}
\end{align}
The interaction of the heavy neutrinos with Higgs has the following form:
\begin{align}
-\mathcal{L}_{H}  = \frac{g M_j}{4 M_W}  H \left\{  (U^\dag_{PMNS} 
V)_{ij}{\bar{\nu}}_i  P_R N_j
  + {\rm H.c.} \right\}
\label{HIGGS}
\end{align}
In the above $P_{L/R} = (1\mp \gamma^5)/2$ is the left/right-chirality 
projection operator,  and $V$ is the mixing 
matrix through which light 
neutrinos mix with the RH neutrinos, and referred to as  active-sterile 
mixing. We consider a diagonal basis for the charged leptons.

For the RH neutrino, coupled with LQ, we do not assume any particular model. 
Instead, we are interested in  different  frameworks of RH neutrinos, 
that can lead to large active-sterile mixing, 
so that the heavy neutrinos decay inside the detector. It is widely known, that 
a number of different frameworks can generate large active-sterile mixing, 
including inverse and linear seesaw~\cite{inverseseesaw1,inverseseesaw2,inverseseesaw3,Linearseesaw1,Linearseesaw2}, 
extended seesaw~\cite{Extendedseesaw1,Extendedseesaw3,Kang:2006sn,Majee:2008mn,Parida:2010wq}, 
cancellation framework \cite{Mitra:2011qr}.   In the inverse seesaw, light SM 
neutrino  masses are extremely tiny, owing  to the small lepton number 
violating parameter of the model. The active-sterile neutrino mixing is not 
constrained from light neutrino masses in this model.  Active-sterile mixing upto $ \mathcal{O}(10^{-2})$ is allowed from  experimental data \cite{heavyNeutrinoMixing1,ATLAS:2012yoa,
heavyNeutrinoMixing3,Khachatryan:2016olu,Sirunyan:2018mtv,Atre:2009rg,Sirunyan:2018xiv,Abada:2007ux}. In extended seesaw, or double 
seesaw~\cite{Chakrabortty:2010az}, the RH neutrino gets mass due to seesaw, and light 
neutrino masses are generated due to two-fold seesaw.  In other 
frameworks, such as, cancellation, small light neutrino masses are  
generated 
due to cancellation between  different RH neutrino contributions in the mass 
matrix \cite{Mitra:2011qr}. The active-sterile mixing is yet unconstrained from 
neutrino data. In all the above mentioned frameworks,   
owing to the charged current and neutral current interactions as well as 
the interaction with the Higgs, specified above in Eqs.(3),(\ref{NC}) 
and 
(\ref{HIGGS}), the RH neutrino $N$ can decay to a number of SM 
particles, 
including  
 $l^{\pm}W^{\mp}$, $\nu Z$, and $\nu H$. 
The branching ratio  of these three decays is $\rm{Br(N\to lW)}: \rm{Br(N\to 
\nu Z)}: \rm{Br(N\to \nu H)} \simeq 0.6:0.3:0.1$, once RH neutrino mass becomes 
larger than the Higgs mass $M_N > M_H$ and $M_{N}<200$~GeV~\cite{Atre:2009rg,Banerjee:2015gca}.

In the following sections, we first consider the  resonant  production of LQ 
and its decay to 
a lepton and jet. We next consider the production of heavy 
neutrinos from LQ decay, and discuss the discovery prospect of the LQ 
in a number of channels. As mentioned before,  we consider the prompt 
decays of heavy neutrino for the analysis of the RH neutrino signature, that 
occurs due to large active-sterile neutrino mixing. We compare 
between the usual charged current (CC) production of heavy neutrinos vs the alternate production from LQ decay. We show that  the production from LQ 
decay dominates over the CC production mode by order of magnitude for active-sterile mixing {$V \lesssim 10^{-2}$. }

\section{Constraints on Leptoquark Couplings}
\label{Constraints on Leptoquark Couplings}
The couplings of the LQs to fermions are constrained by low energy precision observables such as atomic parity violation,
Kaon decays etc. We assume that the Yukawa coupling matrix elements, $Y_{ij}=\delta_{ij}Y_{ii}$ and $Z_{ij}=\delta_{ij}Z_{ii}$, where, $i,j=1,2,3$. Hence the LQ couples exclusively to
a lepton and a quark of the same generation, although it can have non-zero couplings to fermions of more than one generation. 
\begin{itemize}
\item 
{LQs have been searched for and studied in the context of $e^{+}e^{-}$\cite{eecollider1986,eecollider1999,eecollider1999v2,eecollider2001,eecollider2003}, $ep$ \cite{ep1993,SearchForLQatHERA,
ep2003,ep2011,ep2011v2,ep2012},
$p\bar{p}$\cite{ppbar1994,ppbar2008,ppbar2009,ppbar2010,ppbar2011}, and $pp$\cite{pp1993,pp2005,pp2006,pp2008,ppCMS2016,ppATLAS2016} colliders.
The present tightest bounds are from the LHC \cite{ThirdGenerationScalarLQatCMS,SearchForScalarLQatATLAS,CMS:2018bhq,CMS:2018hjx,ATLAS:2017vke}. LHC 
has studied the process $pp\to LQ\,\bar{LQ}
\to\ell j\ell j$ for LQs of first, second and third generations. 
 Non-observation of any new physics at the LHC has ruled out LQs of masses up 
 to 1.1 TeV at 95$\%$ C.L for the LQ decaying
to $e j$ with 100$\%$ branching ratio \cite{SearchForScalarLQatATLAS}. For second generation, the bound is even more stringent $M_{LQ} > 1.5$ TeV at 95$\%$ C.L \cite{ATLAS:2017vke}.
 For third generation, the bound is $M_{LQ} > 900$ GeV at 95$\%$ C.L\cite{CMS:2018hjx} .}
\begin{figure}[h]
\centering
\includegraphics[width=0.45\textwidth]{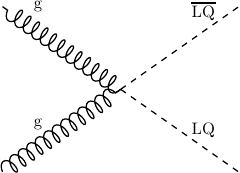}
\hspace{1cm}
\includegraphics[width=0.3\textwidth]{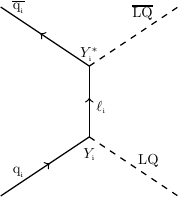}
\caption{\small{Left panel: Feynman diagram for the gluon-initiated LQ pair-production process at LHC. Right panel: the same, but for the quark-initiated processes.}}
\label{feynman diagram for LHC}
\end{figure}
At LHC, numerous QCD diagrams contribute to the LQ pair 
production. For illustration, we show only one representative gluon 
initiated diagram  in the 
left panel of Fig.~\ref{feynman diagram for LHC}.
However, with non-zero Yukawa couplings, significantly large contribution to the LQ pair production may arise through a singlet t-channel diagram (see the right panel of Fig.~\ref{feynman diagram for LHC}).
The  pair production cross-section at LHC  can be parametrised as 
\cite{CorneringScalarLQatLHC}, 
\begin{align}
 \sigma_{\text{pair}}\left(Y_{ii},M_{LQ}\right)=a_{0}(M_{LQ})+a_{2}(M_{LQ})|Y_{ii}|^{2}+a_{4}(M_{LQ})|Y_{ii}|^{4},
\end{align}
where the three terms correspond to the  QCD pair production, { an 
interference term and} t-channel production. In 
Fig.~\ref{LHCpairProduction}, we  show  the variation of LQ pair production 
cross-section
with the Yukawa coupling $Y_{11}$. 

\begin{figure}[h]
\centering
\includegraphics[width=0.5\textwidth]{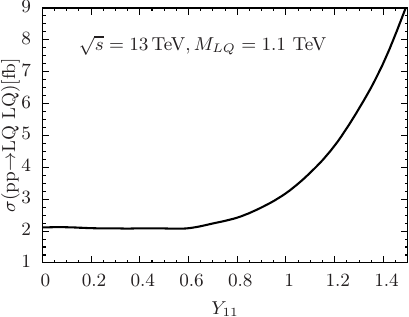}
\includegraphics[width=0.45\textwidth]{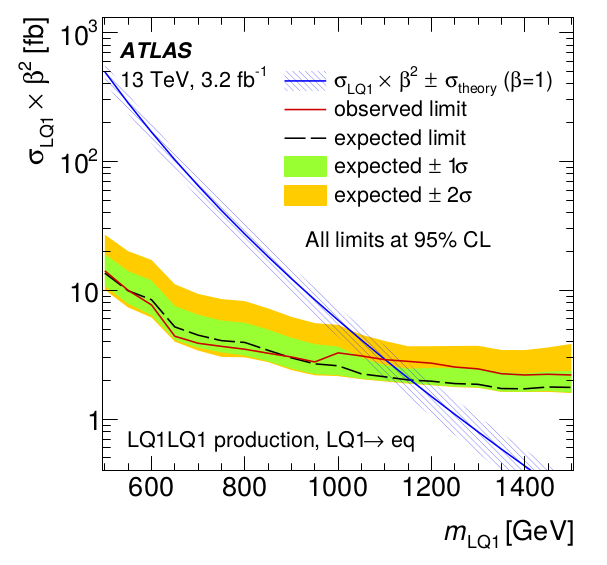}
\caption{\small{Left panel:  variation of the production cross-section with Yukawa for $\sqrt{s}=13$ TeV. Right panel:  limit on scalar LQ pair-production time branching fraction to $eq$ final state as a function of mass for first-generation LQs. The yellow and green bands represent 
the  $2\sigma$ and $1\sigma$ expected limits. The NLO prediction is shown in blue curve with uncertainty due to choice of PDF set and renormalisation/factorisation scale~\cite{SearchForScalarLQatATLAS} }. 
}
\label{LHCpairProduction}
\end{figure}

For small Yukawa coupling $Y_{11}$, LQ pair production is mostly governed by  
QCD, { as can be seen from the straight line upto $Y_{11} \sim 0.5$ in 
Fig.~\ref{LHCpairProduction}. For 
intermediate Yukawa couplings there exists a region with negative interference 
between QCD diagrams and the t-channel diagram where the total cross section 
decreases~\cite{CorneringScalarLQatLHC}, resulting in the mild dip in the 
cross-section for coupling beyond 0.5, that 
is seen in Fig.~\ref{LHCpairProduction}.} 
 { For large Yukawa coupling $Y_{ii}$, the t-channel process dominates 
 and significantly enhances the cross-section.}
 The right panel of {Fig.~\ref{LHCpairProduction} shows the  limit on the first-generation scalar LQ 
 pair-production times the branching fraction to $ej$ final state as a function 
 of the mass. For the branching fraction $LQ \to e j$ as $100\%$, the bound on 
 the pair-production of LQ becomes $\sigma(p p \to LQ LQ) \lesssim 3$ fb, for LQ mass 
 1.1 TeV. Comparing  the left and right panel of Fig.~\ref{LHCpairProduction}, 
 one can see that the limit on the cross section for a 
 LQ of mass 1.1 TeV, will be inconsistent with a 
 Yukawa coupling {larger than} 1.}

It is obvious from Eq.~\ref{eq1}, that for the LQ to have 100$\%$ 
branching ratio in the $LQ \to ej$ decay mode, the coupling $Z$ needs to be 
zero. Allowing  non-zero value for coupling $Z$ will open up new decay modes, 
such as 
$\bar{t} N$ for LQ, and hence will  lower the stringent bound on LQs. We however, 
adopt a conservative approach, and in order to be consistent with the 
LHC results for first generation of LQ, throughout our study, we consider 
$M_{LQ}\geq 1.1$ TeV.  Additionally, we also  keep both the couplings 
non-zero.

\item
{The present bound on coupling $Y$ from atomic parity violation are $Y_{de}<0.34\left(\frac{{M}_{{LQ}}}{1{TeV}}\right)$,
$Y_{ue}<0.36\left(\frac{{M}_{{LQ}}}{1{TeV}}\right)$ \cite{CorneringScalarLQatLHC}.
These bounds are extracted under the assumption that only one of the
two contributions is present at a given moment.  These bounds allow  large coupling for larger mass of LQ, and place a stringent constraint for lighter LQ.}

\item
The most stringent bound on the diagonal couplings of $\tilde{R_2}^{\frac{2}{3}}$ 
comes from LFV decay mode $K_{L}\to\mu^{-}e^{+}$, as this is a tree level process. Following  Refs. \cite{CorneringScalarLQatLHC,LimitonscalarLQ},
the bound is given by, $\left|Y_{s\mu}Y_{de}^{*}\right|<2.1\times 
10^{-5}\left(\frac{{M}_{{LQ}}}{1{TeV}}\right)^2$. In order to 
satisfy both the APV and LFV constraints, for $Y_{de} \sim \mathcal{O}(0.1)$, 
the other 
coupling  $Y_{s\mu}$ has to be tiny.
We consider  $Y_{s\mu}$ to  be zero and a  large value $(0.3)$ for $Y_{de}$ to 
get large production cross-section of LQ at LHeC.

 We discuss the production of a LQ, and  its decay to 
different final states in the next section, for the benchmark points, that are 
in agreement with the described constraints. 
\end{itemize}

\section{Leptoquark Production and Its Decays}
\label{Leptoquark Production}
\begin{figure}[h]
\centering
\includegraphics[width=0.3\textwidth]{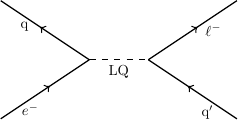}
\hspace{1cm}
\includegraphics[width=0.2\textwidth]{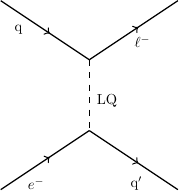}
\caption{\small{Feynman diagram for $e^{-}p\to \ell j$}.}
\label{feynman diagram for eptolj}
\end{figure}

At $e^{-}p$ colliders, scalar LQs can 
be resonantly produced through 
s-channel process as shown in the left panel of Fig.~\ref{feynman diagram for 
eptolj}, and decay to a lepton and a jet. In addition, LQ can also be a  t-channel mediator  for the process $e^{-} p \to l^{-} j$, that we consider in our analysis (shown in the
right panel of Fig.~\ref{feynman diagram for eptolj}).

\begin{figure}[h]
\centering
\includegraphics[width=0.45\textwidth]{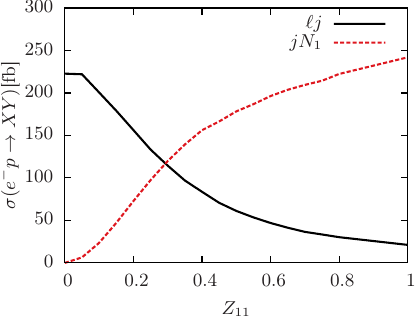}
\includegraphics[width=0.45\textwidth]{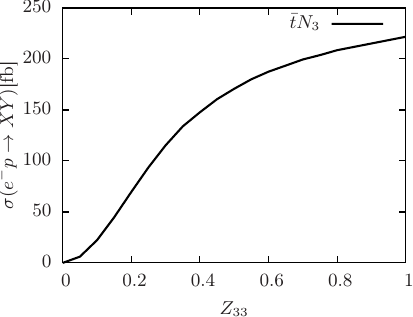}
\caption{\small{Production cross-section for $\ell j$, $jN_{1}$ and $\bar{t}N_{3}$ at LHeC with varying coupling $Z_{11}$. We have considered 150 GeV electron beam colliding with 7 TeV Proton beam. The LQ mass has been set to 1.1 TeV}. For $l^{-}j, j N_1$, the coupling $Y_{11}=0.3$, $Z_{11}$ is varying and for $\bar{t}N_{3}$ production, the coupling $Y_{11}=0.3$, $Z_{33}$ is varying, rest of the Yukawa coupling has been set to zero.}
\label{crossection1}
\end{figure}

{The production cross-section of a LQ at LHeC, as well as that 
for both the single and pair production at LHC, are shown  in 
Fig.~\ref{Production 
crossection} for varying LQ mass. Clearly, the LHeC cross-section is 
more than both the pair-production, as well as, the single production of LQ 
associated with a charged lepton at LHC.} 
The higher LQ production cross-section as well as the lower background 
at LHeC will allow more precise studies for probing LQ and  RH 
neutrinos.  Once produced,  the LQ can decay 
into a number of final states, including,  a) a quark-lepton pair that gives
rise to single  charged lepton and a light jet, b) a light jet and  a heavy 
neutrino,  and c) a top quark accompanied with a heavy neutrino. These heavy 
 neutrinos appearing from the decays of the LQ can again be 
more easily probed at LHeC through its decay products. Note that, for all these processes, the LQ can also mediate as $t$-channel mediator. For  b) and c), there is also $t$-channel contribution from $W$ gauge boson mediator, but significantly smaller for the active-sterile mixing 
$V \lesssim 10^{-2}$. We give  numerical estimates in Section.~\ref{Signals and Background}. However, during computation (in Fig.~\ref{crossection1} and for the collider analysis), we consider all  the contributions together. 

\begin{figure}[h]
\centering
\includegraphics[width=0.45\textwidth]{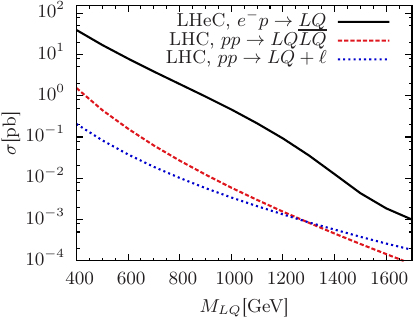}
\caption{\small{Comparison of the
cross section for LQ production at LHC and at LHeC. The c.m.energy for   LHC is $\sqrt{s}= 13$ TeV. For LHeC, we use electron beam  of  150 GeV  and proton beam of  7 TeV, respectively. The coupling $Y_{de}$ has been set to 0.3, in agreement with the experimental constraints.}}
\label{Production crossection}
\end{figure}
\begin{table*}[ht]
	\centering
	\begin{tabular}{|c|c|c|c|c|c|}
		\hline
		Benchmarks & $M_{N_{1,3}}$ & $Y$ & $Z$ & Process & $\sigma_{XY}$(fb)\\
		\hline
		BP1 & $(150,1000,1000)$ & $(0.3,0,0)$  & $(0,0,0)$ & $\ell j$ & $221$  \\
		BP2 & $(150,1000,1000)$ & $(0.3,0,0)$  & $(1,0,0)$ & $jN_{1}$ & $242$  \\
		BP3 & $(1000,1000,150)$ & $(0.3,0,0)$  & $(0,0,1)$ & $\bar{t} N_{3}$ & $222$  \\
		\hline
	\end{tabular}
	\caption{Benchmark parameters and production cross-section for $\ell j$, 
	$jN_{1}$ and $\bar{t}N_{3}$ at LHeC with electron and proton beam energy 
	150 GeV and 7 TeV respectively.
	LQ mass is considered as $1.1\,\text{TeV}$. }
	\label{tab1}
\end{table*}  
For our computations, the LQ mass has been set to 1.1 TeV. We 
choose three benchmark points, with the three heavy neutrino masses and the LQ 
couplings, $Y_{ii}$ and $Z_{ii}$ chosen such that they are consistent 
with all the constraints mentioned in Sec.~\ref{Constraints on Leptoquark Couplings}, as well as with the neutrino oscillation data~\cite{Subhadeep}. These parameters 
for the benchmark points have been specified in Table \ref{tab1}.
The production cross-section for these three processes at LHeC, with electron and proton beam energies of 150 GeV and 7 TeV respectively, are also shown in Fig.~\ref{crossection1} as
a function of the couplings.

The general expression for the two body decay of a scalar LQ to $\ell_{i}q$ and $N_{i}q$ final states are given by,
\begin{align}
 &\Gamma(LQ\to\ell_{i}q)=\frac{|Y_{ii}|^{2}}{16\pi M_{LQ}^{3}}\lambda^{\frac{1}{2}}(M_{LQ}^{2},m_{\ell_{i}}^{2},m_{q^{2}})(M_{LQ}^{2}-m_{\ell_{i}}^{2}-m_{q}^{2})\\ 
 &\Gamma(LQ\to N_{i}q)=\frac{|Z_{ii}|^{2}}{16\pi M_{LQ}^{3}}\lambda^{\frac{1}{2}}(M_{LQ}^{2},M_{N_{i}}^{2},m_{q^{2}})(M_{LQ}^{2}-M_{N_{i}}^{2}-m_{q}^{2})
\end{align}
In the massless limit of leptons and quarks, the branching ratios are given by,
\begin{align}
\label{Branching Ratio}
 \beta(LQ\to\ell_{i}q)=\frac{|Y_{ii}|^{2}}{\sum_{i}(|Y_{ii}|^{2}+|Z_{ii}|^{2})}\,\,\text{and}\,\,\beta(LQ\to N_{i}q)=\frac{|Z_{ii}|^{2}}{\sum_{i}(|Y_{ii}|^{2}+|Z_{ii}|^{2})}
\end{align}

At an $e^{-}p$ collider LQs can be resonantly produced, followed by 
their decay. Hence, we can write the cross section approximately as,
\begin{align}
\sigma (e^{-}p\to \ell_{i}q\,\,\text{or}\,\,N_{i}q)\approx\sigma (e^{-}p\to LQ).\beta
(LQ\to\ell_{i}q\,\,\text{or}\,\,N_{i}q).
\label{cross section times branching ratio}
\end{align}
As can be seen, from Eq.~\ref{Branching Ratio},  with increasing coupling $Z_{11}$ the branching ratio of  $\sigma(e^{-} p \to j N_1)$ 
increases, while $\sigma(e^{-} p \to l j)$ decreases. This results in larger 
cross-section for $\sigma(e^{-} p \to j N_1)$ for larger $Z_{11}$.  Cross 
section for the other channel $e^{-} p \to \bar{t} N_3$ is also large for large value 
of $Z_{33}$. The values of the cross-section in fb, for three benchmark 
points are given in the last column of Table.~\ref{tab1}.  {As can be seen, the 
production cross-section at LHeC is fairly large, approximately $\sigma \sim 
221-242$ fb for the chosen benchmark points. As we will show in the next 
sections, folded with branching ratios of heavy neutrino, top quark, the total cross-section for the 
different final states will be sizeable.}

\begin{figure}[h]
\centering
\includegraphics[width=0.45\textwidth]{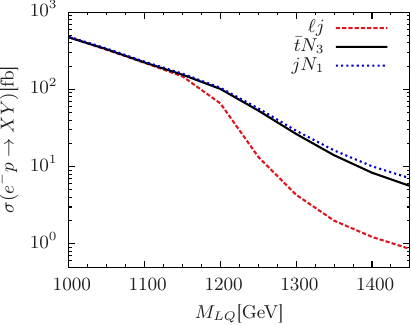}
\caption{\small{Production cross-section for $\ell j$, $jN_{1}$ and $tN_{3}$ with varying LQ mass $M_{LQ}$.  For $l^{-}j, j N_1$ and $\bar{t}N_{3}$ production we adopt BP1, BP2 and BP3 respectively.}}
\label{scanning}
\end{figure}

\section{Collider Analysis}
\label{Collider Analysis}
We implemented the model in FeynRules \cite{FeynRules}, generated the model 
files for \texttt{MadGraph5\_aMC@NLO}\\
(\texttt{v2\_5\_5})~\cite{MadGraph} to calculate the parton level cross-section 
for signals and background.
For the collider simulation part, we passed the MadGraph generated parton level 
events to PYTHIA (v6.4.28) \cite{Pythia6}, where subsequent decay, initial 
state radiation, final state radiation and
hadronisation have been carried out. The jets are reconstructed by anti-$\kappa_{t}$ algorithim \cite{AntiKt} implemented in Fastjet package \cite{FastJet}  with {radius parameter}  $R=0.4$.
For the analysis of signal and background events we use the following set of  basic cuts,\\
\begin{enumerate}
\item
Electrons and muons in the final state should have the following transverse momentum and pseudo-rapidity $p_{T}^{\ell}>20\,\text{GeV}$, $|\eta^{\ell}|<2.5.$
\item
Jets are ordered in $p_{T}$, jets should have $p_{T}^{j}>40\text{GeV}$ and $|\eta^{j}|<2.5$.
\item
Photons are counted if $p_{T}^{\gamma}>10\,\text{GeV}$ and $|\eta^{\gamma}|<2.5$.
\item
Jets should be separated by $\Delta R_{jj}>0.5$.
\item
 Leptons should be separated by $\Delta R_{\ell\ell}>0.2$.
 \item
 Leptons and photons isolation $\Delta R_{\ell\gamma}>0.2$.
\item
 Jets and leptons should be separated by $\Delta R_{\ell j}>0.4$.
\item
 Hadronic activity within a cone of radius $0.3$ around a lepton must be limited to $\sum p_{T}^{\text{hadron}}<0.2 p_{T}^{\ell}$, where $p_{T}^{\ell}$ is the
transverse momentum of lepton within the specified cone.
\end{enumerate}

Due to the  initial and final state radiations,  additional jets {will be present in } the final states considered. 
For the  inverse seesaw framework, lepton number violation (LNV) is dictated by the parameter $\mu_{X}$, that is negligibly small. 
{ Therefore,  the cross-section for LNV di-lepton final states will be suppressed. For the framework where light neutrino masses are generated as a result of cancellation, 
sizable lepton number violation can however be present. Below, we adopt a conservative approach, and only consider lepton number conserving signatures. A number of signatures, including single lepton and
multi-jet, di-lepton associated with multi-jet and missing energy, and multi-lepton associated with missing energy and $b$-jet have been analysed in the subsequent sections. }
\section{Signals and Background}
\label{Signals and Background}
\subsection{Signal I : $\ell^{-}+\text{n-jets}\,(1\leq \text{n} \leq 2)$}
{The single-lepton associated with jet is the easiest channel to probe LQ. LQ, once produced  resonantly, can directly decay to a charged-lepton and a jet.  Additionally, the $t$-channel contribution, as shown in 
Fig.~\ref{feynman diagram for eptolj} will also be present. The    parton-level final state are therefore $\ell^{-}+\text{n-jets}\,(\text{n}=1)$.  Additional jets will be present due to ISR, FSR. We demand the final state should contain $\ell^{-}$ and number of jets $1 \leq \text{n-jets} \leq 2 $.  The main backgrounds arise   from SM  process, such as,  
$e^{-}p\to\ell^{-}j,\,\ell^{-}jj$, that is significantly larger as compared to 
the signal.  From Tab.~\ref{tab2}, the signal cross-section is 220 fb, while 
the background cross-section is ~$3\times 10^{6}$ fb.  We use a number of cuts on different kinematic variables to reduce  the background.}

\begin{figure}[h]
\centering
\includegraphics[width=0.45\textwidth]{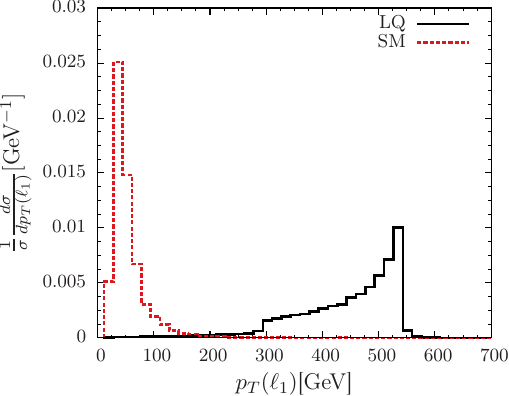}
\includegraphics[width=0.45\textwidth]{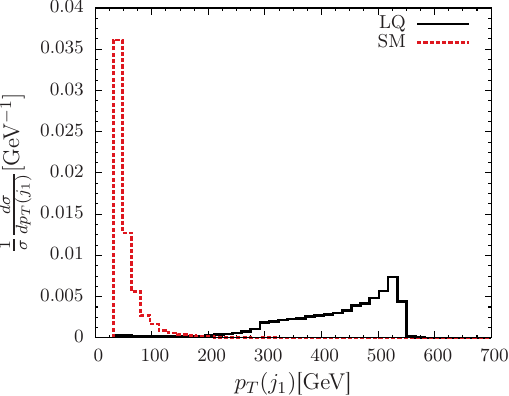}
\includegraphics[width=0.45\textwidth]{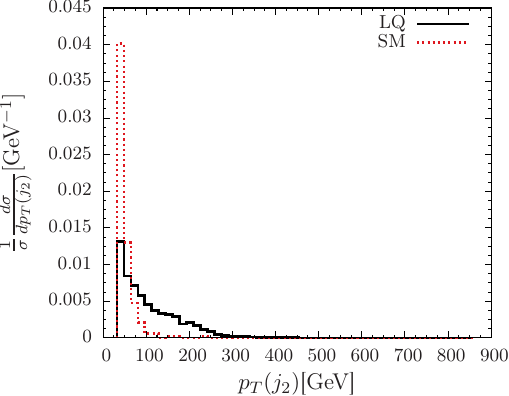}
\includegraphics[width=0.45\textwidth]{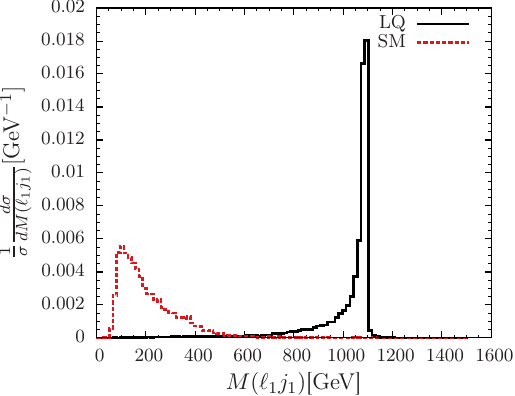}
\caption{\small{ { Distribution of transverse momentum $p_T$ of leading lepton, leading and sub leading jet $p_{T}$ distribution, and invariant mass distribution of leading lepton and leading jet  for the final state
$\ell^{-}+\text{n-jets}\,(1\leq \text{n} \leq 2)$}.}}
\label{distribution for signal 1}
\end{figure}

{In Fig.~\ref{distribution for signal 1} we have shown the transverse 
momentum of the leading lepton, leading and subleading jet, as well as  
the invariant mass distribution of the leading
jet and leading lepton, both for the signal and background.  Evidently, for a very heavy LQ, a  high-$p_{T}$ cut on  leading jet or lepton, and  LQ invariant mass-cut will reduce   SM background drastically.}
\begin{table*}[h]
	\centering
	\begin{tabular}{|c|c|c|c|}
		\hline
		Cuts & Final States & Signal (fb) & Background (fb) \\ 
		\hline
		No cuts & $\ell^{-}+$n-jets$(1\leq n\leq 2)$ & $220$  & $2.96\times 10^{6}$ \\
		Basic cuts & $\ell^{-}+$n-jets$(1\leq n\leq 2)$ & $159$  & $4.08\times 10^{5}$ \\
		Leading lepton $p_{T}$ cut & +$p_{T}^{\ell^{-}}(>400\text{GeV})$ & $118$  & $178$ \\ 
		LQ invariant mass cut & +$|M_{LQ}-M_{\ell_{1}j_{1}}|\leq 100\text{GeV}$ & $101$  & $0$\\ 
		\hline
	\end{tabular}
	\caption{{Signal and Background cross-sections for the final state $\ell^{-}+\text{n-jets}\,(1\leq \text{n} \leq 2)$ after different cuts. BP1 has been used for this final state.}}
	\label{tab2}
\end{table*}  
In Table.~\ref{tab2}, we have shown how we can reduce the SM background to zero 
using  $p_{T}$ cut - $p_{T}^{\ell^{-}}(>400\text{GeV})$ on leading lepton and 
invariant mass cut - 
$|M_{LQ}-M_{\ell_{1}j_{1}}|\leq 100\text{GeV}$ simultaneously after using the basic sets of cuts mentioned in the previous section.
\subsection{Signal II}
\label{Signal II}
\begin{figure}[h]
\centering
\includegraphics[width=0.4\textwidth]{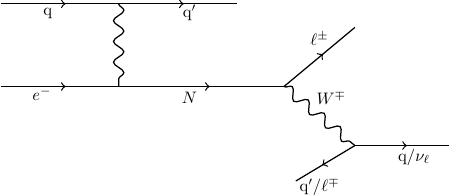}
\hspace{1cm}
\includegraphics[width=0.4\textwidth]{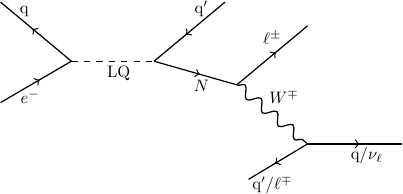}
\caption{\small{Feynman diagram for various final states from $e^{-}p\to jN$. For this case, t-channel LQ mediated diagram can also contribute}.}
\label{feynman diagram for eptojn}
\end{figure}

{ If the coupling $Z_{ij}$ is non-zero, the LQ can also decay to RH neutrino and a jet, as shown in   the right panel of Fig.~\ref{feynman diagram for eptojn}.
The considered final state, can also arise from  the  t-channel $W$ exchange diagram as shown in left panel of Fig.~\ref{feynman diagram for eptojn}.
For active-sterile mixing $V \sim 10^{-2}-10^{-3}$,  the  contribution from LQ  
however dominates. For example, with the BP2, the CC production cross section is $\approx 12.7$~fb, while the production cross section from LQ decay is $\approx 240$~fb.} 
 The subsequent decays of RH neutrino, followed by hadronic and leptonic decays of gauge bosons gives rise to a number of partonic states, that we list below. 
 \begin{enumerate}
 \item $\ell^{-}+\text{n-jets}\,(\text{n}=3)$ (For hadronic decays of $W^{+}$)\\
\item $\ell^{-}+\ell^{+}+\text{n-jets}\,(\text{n}=1)+\slashed{E}_{T}$ (For leptonic decays of $W^{+}$)
\end{enumerate}
\subsubsection{$\ell^{-}+\text{n-jets}\,(\text{n}\geq 3)$}
\begin{figure}[h]
\centering
\includegraphics[width=0.45\textwidth]{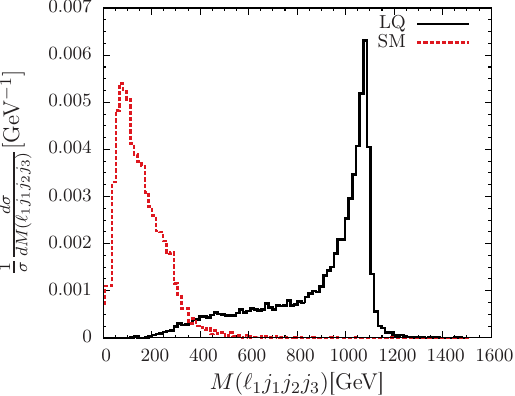}
\includegraphics[width=0.45\textwidth]{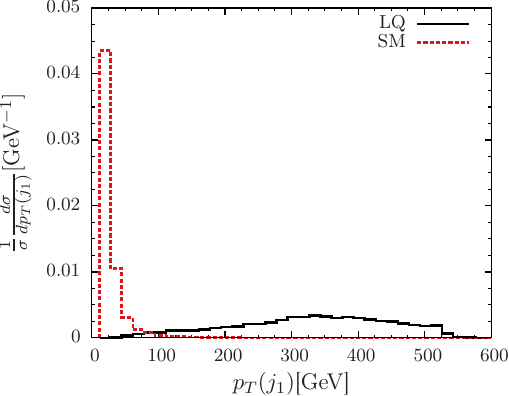}
\includegraphics[width=0.45\textwidth]{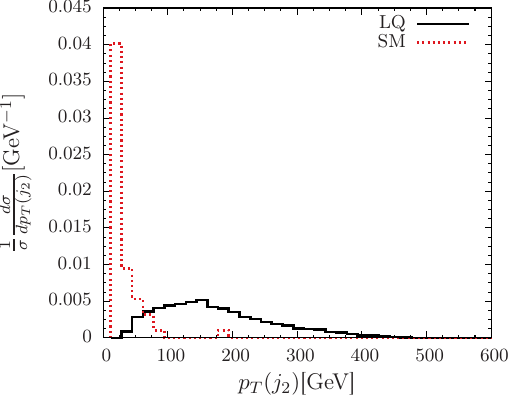}
\includegraphics[width=0.45\textwidth]{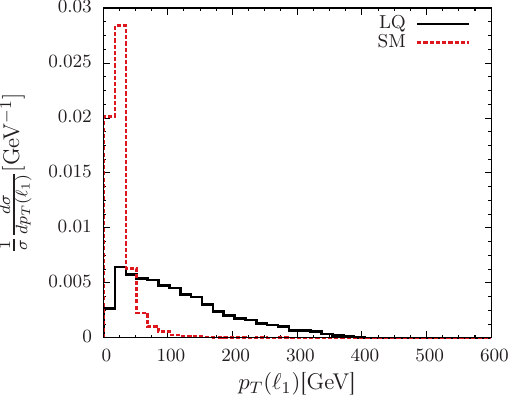}
\includegraphics[width=0.45\textwidth]{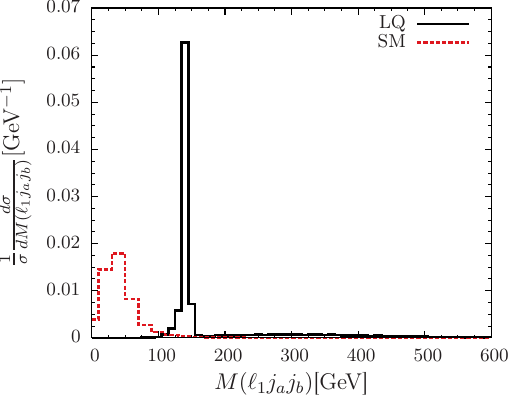}
\caption{\small{Invariant Mass distribution of LQ,  $p_{T}$ distribution of leading, sub leading jet, and of leading lepton. We also show the  invariant mass distribution of $N$ for the final state
$\ell^{-}+\text{n-jets}\,(\text{n}\geq 3)$.}}
\label{distribution for signal 2.1}
\end{figure}

For the case of the hadronic decays of the charged gauge boson,
we demand a charged lepton and at least three jets as final state. The invariant mass of the three jets and the charged lepton must be equal to that of the {mass of the  LQ}.
Hence { a cut on the }  invariant mass distribution allows the separation of the signal from the backgound. For the background we generate  $e^{-} p \to l^{-} + \textrm{n-jets}$ upto $n=3$.  The distribution is given in the  top left panel of Fig.~\ref{distribution for signal 2.1}. {Additionally, the 
leading jet that is directly generated from LQ decay has a very high transverse momentum (see Fig.~\ref{distribution for signal 2.1}). Therefore, a  large cut on the {transverse momentum  of the} leading jet reduces the background}. As can be seen from Table.~\ref{tab3}, { the large $p_T$ cut on the leading jet itself reduces the background by $\mathcal{O}(10^{4})$.  Further reduction in background is achieved though a cut on the invariant mass distribution of the heavy neutrino.
Stringent cuts, such as, the cuts on the invariant mass of the LQ and the heavy neutrino make the background negligibly small.} 
\begin{table*}[h]
	\centering
	\begin{tabular}{|c|c|c|c|}
		\hline
		Cuts & Final States & Signal (fb) & Background (fb) \\ 
		\hline
		No cuts & $\ell^{-}+$n-jets$(n\geq 3)$ & $24.8$  & $2.99\times 10^{6}$\\ 
		Basic cuts & $\ell^{-}+$n-jets$(n\geq 3)$ & $7.65$  & $2.9\times 10^{3}$ \\ 
		Leading Jet $p_{T}$ cut & +$p_{T}^{j_{1}}(>200\text{GeV})$ & $6.56$  & $180$\\ 
		LQ invaraint mass cut & +$|M_{LQ}-M_{\ell_{1}j_{1}j_{2}j_{3}}|\leq 100\text{GeV}$ & $3.65$  & $60$ \\ 
		$N$ invaraint mass cut & +$|M_{N_{1}}-M_{\ell_{1}j_{a}j_{b}}|\leq 30\text{GeV}$ & $3.08$  & $0$ \\ 
		\hline     
	\end{tabular}
		\caption{Signal and Background cross-sections  for the final state $\ell^{-}+\text{n-jets}\,(\text{n}\geq 3)$ with cuts. BP2 has been used for this final state.}
	\label{tab3}
\end{table*}
\subsubsection{$\ell^{-}\ell^{+}+\text{n-jets}\,(\text{n}\geq 1)+\slashed{E}_{T}$}
{For the scenario, when charged gauge boson produced in the decay of RH neutrino, decays leptonically,  the signal will have 2 opposite sign charged leptons, jets (one or more) and missing energy. The dominant SM background comes from the processes like $e^{-}p\to\ell^{-}\ell^{+}j\nu_{\ell}$ and 
$e^{-}p\to\ell^{-}\ell^{+}jj\nu_{\ell}$.
The reduction of the background in this case is achieved through a cut on the  missing energy  and the  cut on effective mass $M_{EFF}$. 
The $M_{EFF}$ variable is   defined as,}
\begin{align*}
 M_{EFF}=\sum_{i}p_{T_{i}}^{j}+\sum_{i}p_{T_{i}}^{\ell}+\slashed{E}_{T}, 
\end{align*}
{where $p_{T_{i}}^{j}$, $p_{T_{i}}^{\ell}$ are the transverse momentum of the jet and lepton, and $\slashed{E}_{T}$ is the missing transverse energy. 
We expect a hard distribution for $M_{EFF}$, since the $p_{T}$ of the lepton and jets coming mostly from resonantly produced LQ (as the t-channel contribution is small) is significantly large. The peak of the $M_{EFF}$ shifts towards higher values with increasing LQ mass.}
\begin{figure}[h]
\centering
\includegraphics[width=0.45\textwidth]{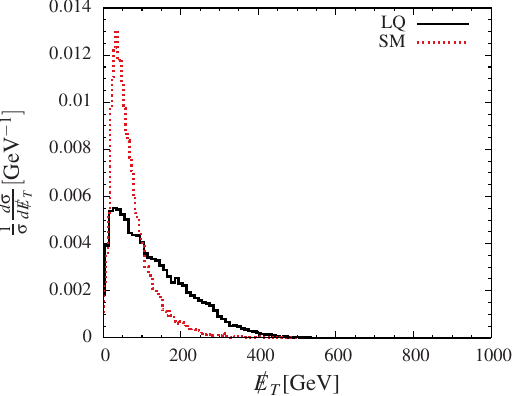}
\includegraphics[width=0.45\textwidth]{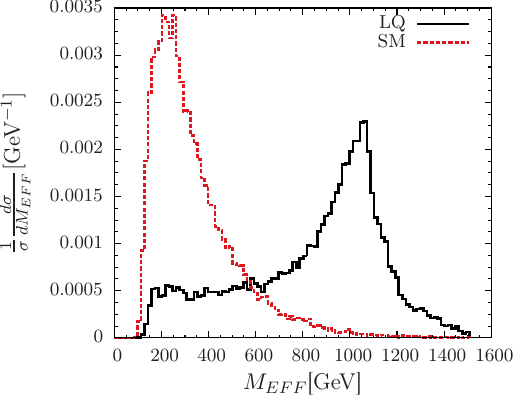}
\includegraphics[width=0.45\textwidth]{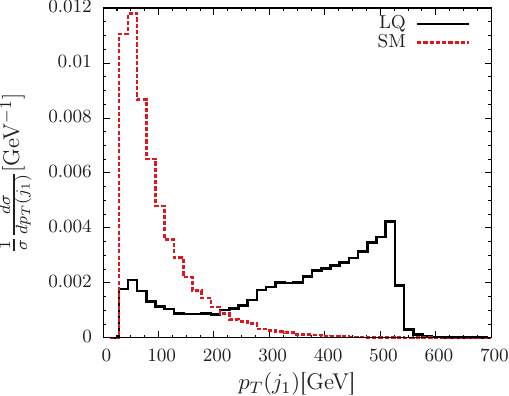}
\includegraphics[width=0.45\textwidth]{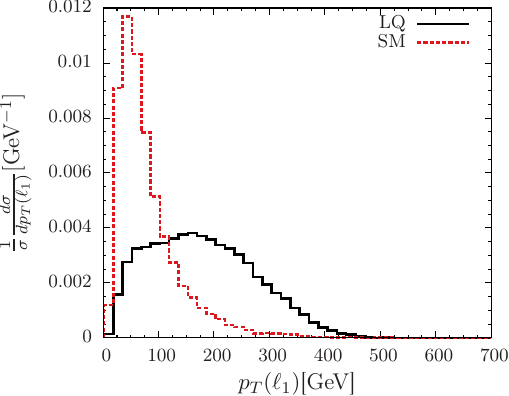}
\caption{\small{Missing transverse energy distribution, $M_{EFF}$ distribution, $p_{T}$ distribution of  leading jet,  and  leading lepton for the final
state $\ell^{-}\ell^{+}+\text{n-jets}\,(\text{n}\geq 1)+\slashed{E}_{T}$.}}
\label{distribution for signal 2.2}
\end{figure}
We have shown in Fig.~\ref{distribution for signal 2.2} the missing energy $\slashed{E}_{T}$, $M_{EFF}$, leading jet and leading lepton $p_{T}$ distributions. 
The effect of different cuts on the signal and background cross-sections are given in Table.~\ref{tab4}.

\begin{table*}[ht]
	\centering
	\begin{tabular}{|c|c|c|c|}
		\hline
		Cuts & Final States & Signal (fb) & Background (fb)\\ 
		\hline
		No cuts & $\ell^{-}\ell^{+}+$n-jets$(n \geq 1)+\slashed{E}_{T}$ & $11.2$  & $5.22\times 10^{2}$ \\ 
		Basic cuts & $\ell^{-}\ell^{+}+$n-jets$(n \geq 1)+\slashed{E}_{T}$ & 7.84  & 258 \\ 
		Missing energy cut & $\ell^{-}\ell^{+}+$n-jets$(n \geq 1)+\slashed{E}_{T}(>100\text{GeV})$ & 4.26  & 57.5 \\ 
		Leading Jet $p_{T}$ cut & +$p_{T}^{j_{1}}(>300\text{GeV})$ & 3.24  & 3.73 \\ 
		$M_{EFF}$ cuts & +$M_{EFF}(>500\text{GeV})$ & 2.88  & 2.54 \\ 
		\hline
	\end{tabular}
	\caption{Signal and Background cross-section after various cuts for the final state $\ell^{-}\ell^{+}+\text{n-jets}\,(\text{n}\geq 1)+\slashed{E}_{T}$. BP2 has been used for this final state. }
	\label{tab4}
\end{table*}  
\subsection{Signal III}
For the LQ mass more than $M_N+M_t$,  it can  further decay to $\bar{t}N_{3}$, 
that enables a final state  with large lepton or large jet multiplicity. 
The large lepton multiplicity is promising due to suppressed  
 SM background. 
\begin{figure}[h]
\centering
\includegraphics[width=0.7\textwidth]{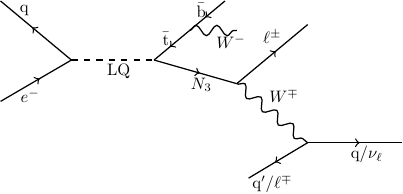}
\caption{\small{Feynman diagram for various final states from $e^{-}p\to \bar{t}N_{3}$. For this case also, the t-channel mediated diagram with gauge boson and LQ will contribute. }}
\label{feynman diagram for eptotn3}
\end{figure}
 For the $\bar{t}N_{3}$ production channel, considering subsequent decays of 
 $N_{3}$ and $\bar{t}$, where $N_{3}$ is assumed to decay to $\tau^\pm W^{\mp}$,  
 following final states at parton-level are 
 possible:\\
(1) $\bar{b}\ell^{-}\tau^{-}\ell^{+}+\slashed{E}_{T}$ (For leptonic decays of both the $W$ bosons, $W^{-}\to\ell^{-}\nu$, $W^{+}\to\ell^{+}\nu$),\\
(2) $\bar{b}\ell^{-}\tau^{-}+\text{n-jets}\,(\text{n}=2)+\slashed{E}_{T}$(For the $W$ boson decays, $W^{-}\to\ell^{-}\nu$ and $W^{+}\to j j$),\\
(3) $\bar{b}\tau^{-}\ell^{+}+\text{n-jets}\,(\text{n}=2)+\slashed{E}_{T}$(For the $W$ boson decays, $W^{-}\to j j$ and $W^{+}\to\ell^{+}\nu$),\\
(4) $\bar{b}\tau^{-}+\text{n-jets}\,(\text{n}=4)$(For the $W$ boson decays, $W^{-}\to j j$ and $W^{+}\to j j$).\\
We do not consider the last final states in our study because of very small cross-section and very large SM background due to large jet multiplicity. 

\subsubsection{$\bar{b}\ell^{-}\tau^{-}\ell^{+}+\slashed{E}_{T}$}
For the final states involving $\tau$ and b, tagging can reduce the SM background significantly. We consider the $p_{T}$ for the b and $\tau$ jets, as $p_{T}> 40$~GeV. In this work, we adopt a minimalistic approach and consider a flat $75\%$ efficiency for b-tagging and $60\%$ efficiency for $\tau$-tagging.
 Similar to the previous case, the $M_{EFF}$ distribution
is hard due  to the  large missing energy and large transverse momenta of final state particles. For this signal, the most dominant SM backgound comes from the
process $\bar{t}\ell^{-}W^{+}$, $\bar{t}Z\nu$ and $\bar{t}h\nu$ with $\bar{t}\to\bar{b}\tau^{-}\nu$, $W^{+}\to\ell^{+}\nu$, $Z\to\ell^{+}\ell^{-}$, and $h\to\ell^{+}\ell^{-}$. After applying the basic cuts only, SM background
drops significantly. In addition, we use missing energy $\slashed{E}_{T}$, leading jet $p_{T}$ and $M_{EFF}$ distribution to further reduce the SM background.
\begin{table*}[ht]
	\centering
	\begin{tabular}{|c|c|c|c|}
		\hline
		Cuts & Final States & Signal (fb) & Background (fb) \\ 
		\hline
		No cuts & $\bar{b}\ell^{-}\tau^{-}\ell^{+}+\slashed{E}_{T}$ & $1.57$  & $0.323$ \\ 
		Basic cuts & $\bar{b}\ell^{-}\tau^{-}\ell^{+}+\slashed{E}_{T}$ & $0.83$  & $7.51\times 10^{-3}$ \\ 
		Missing energy cut & $\bar{b}\ell^{-}\tau^{-}\ell^{+}+\slashed{E}_{T}(>100\text{GeV})$ & $0.502$  & $4.46\times 10^{-3}$ \\ 
		Leading Jet $p_{T}$ cut & +$p_{T}^{j_{1}}(>100\text{GeV})$ & $0.476$  & $2.16\times 10^{-3}$ \\ 
		$M_{EFF}$ cuts, b and $\tau$ tagging & +$M_{EFF}(>500\text{GeV})$ & $0.21$  & $7.7\times 10^{-4}$ \\ 
		\hline         
	\end{tabular}
	\caption{Signal and Background cross-section after various cuts for the final state $\bar{b}\ell^{-}\tau^{-}\ell^{+}+\slashed{E}_{T}$. BP3 has been used for this final state.}
	\label{tab5}
\end{table*}  

\subsubsection{$\bar{b}\ell^{-}\tau^{-}+\text{n-jets}\,(\text{n}\geq 2)+\slashed{E}_{T}$}
For this case, due to large jet multiplicity, SM background is greater than the previous signal and the background mainly comes from the process $\bar{t}\ell^{-}W^{+}$. { However,  using missing energy $\slashed{E}_{T}$, cuts on leading jet $p_{T}$  and
$M_{EFF}$ distribution, the SM background can be reduced significantly. } 
\begin{table*}[ht]
	\centering
	\begin{tabular}{|c|c|c|c|}
		\hline
		Cuts & Final States & Signal (fb) & Background (fb)\\ 
		\hline
		No cuts & $\bar{b}\ell^{-}\tau^{-}+\text{n-jets}\,(\text{n}\geq 2)+\slashed{E}_{T}$ & $3.54$  & $0.729$ \\ 
		Basic cuts & $\bar{b}\ell^{-}\tau^{-}+\text{n-jets}\,(\text{n}\geq 2)+\slashed{E}_{T}$ & $1.457$  & $2.18\times 10^{-2}$  \\ 
		Missing energy cut & $\bar{b}\ell^{-}\tau^{-}+\text{n-jets}\,(\text{n}\geq 2)+\slashed{E}_{T}(>100\text{GeV})$ & $1.277$  & $1.344\times 10^{-2}$ \\ 
		Leading Jet $p_{T}$ cut & +$p_{T}^{j_{1}}(>100\text{GeV})$ & $1.226$  & $7.65\times 10^{-3}$ \\ 
		$M_{EFF}$ cuts, b and $\tau$ tagging & +$M_{EFF}(>500\text{GeV})$ & $0.522$  & $2.4\times 10^{-3}$ \\ 
		\hline         
	\end{tabular}
	\caption{Signal and Background cross-section after various cuts for the final state $\bar{b}\ell^{-}\tau^{-}+\text{n-jets}\,(\text{n}\geq 2)+\slashed{E}_{T}$. BP3 has been used for this final state.}
	\label{tab6}
\end{table*}  

\subsubsection{$\bar{b}\ell^{+}\tau^{-}+\text{n-jets}\,(\text{n}\geq 2)+\slashed{E}_{T}$}
For this final state, SM background is actualy negligibly small at $e^{-}p$ 
collider for the  beam energies considered.
\begin{table*}[ht]
	\centering
	\begin{tabular}{|c|c|c|c|}
		\hline
		Cuts & Final States & Signal (fb) & Background (fb) \\ 
		\hline
		No cuts & $\bar{b}\ell^{+}\tau^{-}+\text{n-jets}\,(\text{n}\geq 2)+\slashed{E}_{T}$ & $4.07$  & $0$ \\ 
		Basic cuts, b and $\tau$ tagging & $\bar{b}\ell^{+}\tau^{-}+\text{n-jets}\,(\text{n}\geq 2)+\slashed{E}_{T}$ & $1.83$  & $0$  \\ 
		\hline         
	\end{tabular}
	\caption{Signal and Background cross-section after various cuts for the final state $\bar{b}\ell^{+}\tau^{-}+\text{n-jets}\,(\text{n}\geq 2)+\slashed{E}_{T}$. BP3 has been used for this final state.}
	\label{tab7}
\end{table*}  

\section{Signal Strength for Higher LQ Mass}
\label{Signal Strength for Higher LQ Mass}
Bound on LQs parameter space is expected to improve in future with increasing luminosity  at LHC. Hence, we repeat our study for higher LQ masses. 
Though for higher LQ mass, we can allow for large yukawa coupling, we use the same coupling as for 1.1 TeV LQ mass to compare our result for different LQ masses.
Using the same set of cuts for each final states as we did for LQ mass 1.1 TeV, we have calculated the cross-section for LQ mass of $1.2,1.3,...\text{upto}\,1.7$ TeV.   
The partonic cross-sections and the effect of different cuts is shown in Table.~\ref{tab8} and~\ref{tab9}. 
\subsection{Zero Background Case}
\begin{figure}[h]
\centering
\includegraphics[width=0.45\textwidth]{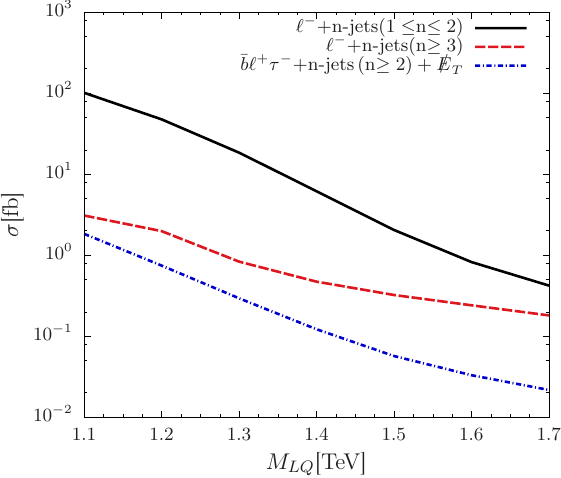}
\includegraphics[width=0.45\textwidth]{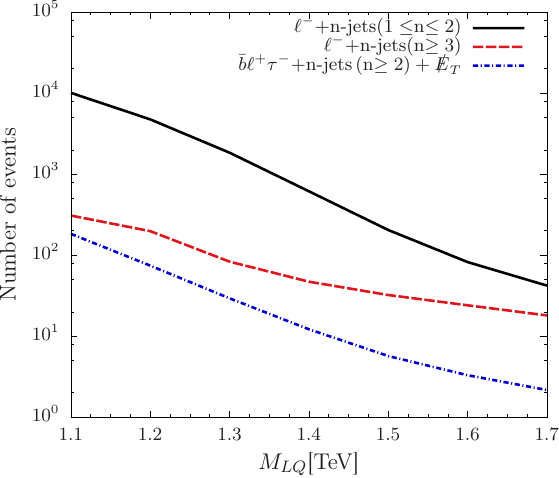}
\caption{Left panel: Signal cross-section as a function of LQ mass for different final states. Right panel: No of events with integrated luminosity $100\,\text{fb}^{-1}$.}
\label{CS and no of events}
\end{figure}

First, we consider only final states for which the SM background is zero or can be reduced to zero using invariant mass cut of LQ and RH neutrinos. The SM background is practically zero, for final state,
$\bar{b}\ell^{+}\tau^{-}+\text{n-jets}\,(\text{n}\geq 2)+\slashed{E}_{T}$. For the final state, $\ell^{-}+$n-jets$(1\leq \text{n}\leq 2)$, 
using $p_{T}^{\ell}$ cut and LQ invariant mass cut (for corresponding LQ mass), the SM background can be reduced to zero. 
\begin{table*}[ht]
	\centering
	\begin{tabular}{|c|c|c|c|c|c|c|c|}
		\hline
		Final States & 1.1 TeV & 1.2 TeV & 1.3 TeV & 1.4 TeV & 1.5 TeV & 1.6 TeV & 1.7 TeV\\
		\hline
		$\ell^{-}+$n-jets$(1\leq \text{n}\leq 2)$ & 101 & 47.53 & 18.37 & 6.148 & 2.04 & 0.82 & 0.42 \\
		$\ell^{-}+$n-jets$(\text{n}\geq 3)$  & 3.08 & 1.98 & 0.83 & 0.47 & 0.322 & 0.24 & 0.18 \\
		$\bar{b}\ell^{+}\tau^{-}+\text{n-jets}\,(\text{n}\geq 2)+\slashed{E}_{T}$ & 1.83& 0.74 & 0.29 & 0.12 & 0.06 & 0.03 & 0.02 \\
		\hline         
	\end{tabular}
		\caption{Cross-sections (in fb) after all the cuts as a function of LQ mass. }
	\label{tab8}
\end{table*}  
Similarly, for the final state, {$\ell^{-}+$n-jets$(\text{n}\geq 3)$, selection cut on  $p_{T}^{j}$, along with on LQ invariant mass, and  invariant mass of RH neutrino can make  the  SM background negligibly small.
The results are given in Table.~\ref{tab8}.} We have also shown the cross section and number of events with $100\,\text{fb}^{-1}$ integrated luminosity as a function of LQ masses in Fig.~\ref{CS and no of events}. {As can be seen, the cross-section for the $\ell^{-}+$n-jets$(1\leq \text{n}\leq 2)$ channel is the largest, varies $10^2-0.42$ fb for a wide range of LQ mass.  With $100\, \rm{fb}^{-1}$ luminosity, this predicts $10^4$ number of events at LHeC. The other channel  with jet multiplicity $(\text{n}\geq 3)$ also offers a  large cross-section, and large number of events  $\mathcal{O}(10^2) $. } 

\subsection{Non-Zero Backgound Case}
For the final states, $\ell^{-}+$n-jets$(1\leq \text{n}\leq 2)$ and $\ell^{-}+$n-jets$(\text{n}\geq 3)$, the SM background is non zero if we do not use the invariant mass of LQ and RH neutrino.
{Since, the LQ  and RH neutrino masses  are unknown, we do not implement the 
mass cut, rather in this section show the cross-sections with a very generic 
sets of cuts. Assuming  LQ mass to be more than 1 TeV, 
all the other cuts which we considered can be
easily applied. For the above  two final states we applied cuts only on  $p_{T}^{\ell}$ and $p_{T}^{j}$.} For final states, $\ell^{-}\ell^{+}+$n-jets$(\text{n} \geq 1)+\slashed{E}_{T}$,
$\bar{b}\ell^{-}\tau^{-}\ell^{+}+\slashed{E}_{T}$ and $\bar{b}\ell^{-}\tau^{-}+\text{n-jets}\,(\text{n}\geq 2)+\slashed{E}_{T}$ we used the same cuts as in Tables.~\ref{tab4}, \ref{tab5} and
\ref{tab6} respectively.
\begin{table*}[ht]
	\centering
	\begin{tabular}{|c|c|c|c|c|c|c|c|}
		\hline
		Final States & 1.1 TeV & 1.2 TeV & 1.3 TeV & 1.4 TeV & 1.5 TeV & 1.6 TeV & 1.7 TeV\\
		\hline
		$\ell^{-}+$n-jets$(1\leq \text{n}\leq 2)$ & 118 & 57 & 23 & 8 & 2.8 & 1.16 & 0.62 \\
		$\ell^{-}+$n-jets$(\text{n}\geq 3)$  & 6.56 & 2.36 & 1 & 0.5 & 0.32 & 0.23 & 0.18 \\
		$\ell^{-}\ell^{+}+$n-jets$(\text{n} \geq 1)+\slashed{E}_{T}$  & 2.88 & 1.38 & 0.57 & 0.23 & 0.102 & 0.057 & 0.039 \\
		$\bar{b}\ell^{-}\tau^{-}\ell^{+}+\slashed{E}_{T}$ & 0.21 & 0.08 & 0.03 & 0.012 & 0.007 & 0.003 & 0.0018 \\
		$\bar{b}\ell^{-}\tau^{-}+\text{n-jets}\,(\text{n}\geq 2)+\slashed{E}_{T}$ & 0.522 & 0.175 & 0.063 & 0.025 & 0.013 & 0.008 & 0.006 \\
		\hline         
	\end{tabular}
		\caption{Cross-sections in fb, after all the cuts (except 
		invariant LQ and right handed neutrinos mass cut) as a function of LQ 
		mass. Backgrounds are same as for 1.1 TeV, as
		the beam energies are same. 
		 }
	\label{tab9}
\end{table*}  
\begin{figure}[h]
\centering
\includegraphics[width=0.45\textwidth]{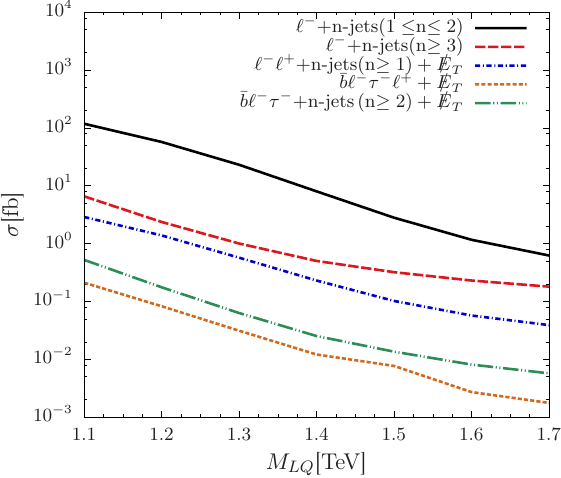}
\includegraphics[width=0.45\textwidth]{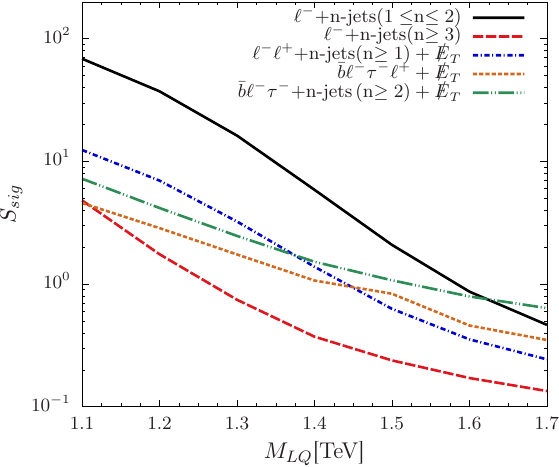}
\caption{\small{Left panel:  The signal cross-section  as a function of LQ mass for different final states. Right panel: the significance with $100\,\rm{fb}^{-1}$ luminosity.}}
\label{cross-section for different final states}
\end{figure}
We  show the signal cross section and statistical significance with integrated luminosity of $100\,\text{fb}^{-1}$ in Fig.~\ref{cross-section for different final states}.
We  also show the required luminosity to achieve $3\sigma$ and $5\sigma$ statistical significance in Fig.~\ref{three and five sigma significance}.
 The statistical significance has been calculated using the following expression:
 \begin{equation}
 {S_{sig}=\frac{S}{\sqrt{S+B}},}
  \end{equation}
 {where, $S$ and $B$ denote the number of signal and background events, respectively.}

\subsubsection{Results}

{{ We discuss  the discovery prospect of LQ in the mass range 1.1-1.7 TeV  at  LHeC. The channel $\ell^{-}+\text{n-jets}(1\leq \text{n}\leq 2)$ is  the most promising.}
For the final states $\ell^{-}+\text{n-jets}(1\leq \text{n}\leq 2)$, even with  integrated luminosity $2\,\text{fb}^{-1}$, the statistical significance 
is $9.69\sigma$ for the LQ of mass 1.1 TeV. To probe larger masses, higher luminosity is required. Assuming $\mathcal{L}=100 \, \rm{fb}^{-1}$,  a 1.7 TeV LQ can be probed at  $0.4\sigma$. 
For this final state, $5\sigma$ statistical significance can be probed for 
LQ mass range~$[1.1-1.4]$, with integrated luminosity of $100\,\text{fb}^{-1}$.

For final state $\ell^{-}+\text{n-jets}(n\geq 3)$, with integrated luminosity $100\,\text{fb}^{-1}$, the statistical significance 
is $4.8\sigma$ for LQ mass of 1.1 TeV, decreasing to $0.13\sigma$ for a 1.7 TeV LQ. 


\begin{figure}[h]
\centering
\includegraphics[width=0.45\textwidth]{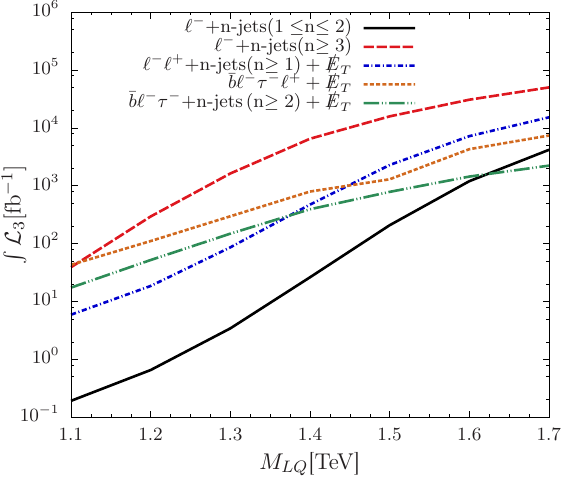}
\includegraphics[width=0.45\textwidth]{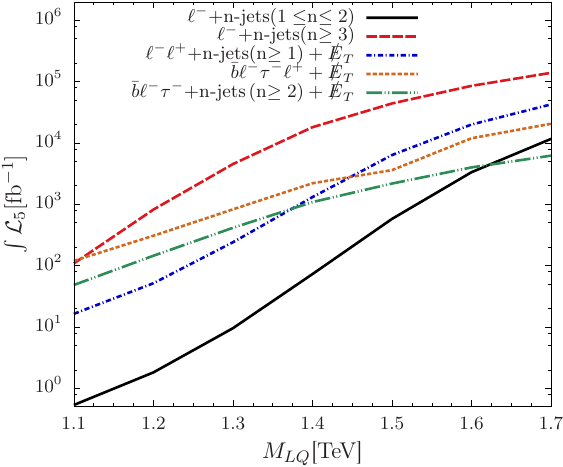}
\caption{\small{The required luminosity to achieve $3\sigma$  (left panel) and $5\sigma$ (right panel ) for different final states.}} 
\label{three and five sigma significance}
\end{figure}

For the final state $\ell^{-}\ell^{+}+\text{n-jets}(\text{n}\geq 1)+\slashed{E}_{T}$, with integrated luminosity $100\,\text{fb}^{-1}$, the statistical significance
is $12\sigma$ for a 1.1 TeV LQ, which decreases to $0.24\sigma$ for a 1.7TeV LQ. For this final state, $5\sigma$ statistical significance can be probed only for 
LQ mass range~$[1.1-1.2]\,\text{TeV}$, with integrated
luminosity $100\,\text{fb}^{-1}$. 

For the final state $\bar{b}\ell^{-}\tau^{-}\ell^{+}+\slashed{E}_{T}$, with integrated luminosity $100\,\text{fb}^{-1}$, the statistical significance
is $4.5\sigma$ for a 1.1 TeV LQ, which decreases to $0.3\sigma$ for a 1.7~TeV LQ. For this final state, $5\sigma$ statistical significance can be probed for LQ mass 
$1.1\,\text{TeV}$, with integrated
luminosity $120\,\text{fb}^{-1}$. In spite of small SM background, as the signal cross section is itself small for higher LQ masses, its difficult to observe this final state for higher LQ masses.

For the final state $\bar{b}\ell^{-}\tau^{-}+\text{n-jets}\,(\text{n}\geq 2)+\slashed{E}_{T}$, with integrated luminosity $100\,\text{fb}^{-1}$, the statistical significance 
is $7\sigma$ for a 1.1 TeV LQ, which decreases to $0.6\sigma$ for a 1.7~TeV LQ. For this final state, $5\sigma$ statistical significance can be probed only for LQ mass 
$1.1\,\text{TeV}$, with integrated
luminosity $100\,\text{fb}^{-1}$. Again for this case also, signal cross section is small for higher LQ mass, hence difficult to probe higher mass regime.
However, it may be noted that our estimates for the cross-sections for LQs of higher masses are rather conservative, as they have been computed, assuming the coupling to be the same as that for 1.1~TeV LQ, while higher values of couplings for larger LQ masses will be permissible.


\section{Conclusions}
\label{Conclusions}
In this work, we study the discovery prospect of  $\tilde{R}_2$ class of LQ model   at LHeC. The model contains two LQs with $Q=\frac{2}{3}$, and  $Q=-\frac{1}{3}$. LQ with $Q=\frac{2}{3}$ can be copiously produced at LHeC, 
due to its interaction with the electron and down type quark. We study the   production  and its  decay to different final states, including a lepton and a jet,   a jet and a  RH neutrino, and RH neutrino and a top quark.  The typical production cross-section for $e^{-}p \to l j, j N_1, \bar{t} N_3 $ are 221, 242, 222 fb for $M_{LQ}=1.1$ TeV, $M_{N_{1,3}}=150$ GeV, and the couplings 
$Y_{11}=0.3, Z_{33}=1$.  The produced RH neutrino further decays and give a plethora of model signatures.  
For the RH neutrinos, we adopt a model independent framework, and a large active-sterile mixing to ensure its  decay within the detector. For the LQs, the higher production cross-section as well as the lower 
backgrounds at LHeC  result in a much higher statistical significance for few of the signals studied. 

We have analysed a number of final states, including $\ell^{-}+\text{n-jets}\,(1\leq \text{n} \leq 2)$, $\ell^{\pm}+\text{n-jets}\,(\text{n}\geq3)$, $\ell^{\pm}\ell^{\mp}+\text{n-jets}\,(\text{n}\geq1)+\slashed{E}_{T}$, $\bar{b}\ell^{-}\tau^{-}\ell^{+}+\slashed{E}_{T}$, $\bar{b}\ell^{-}\tau^{-}+\text{n-jets}\,(\text{n}\geq 2)+\slashed{E}_{T}$, $\bar{b}\ell^{+}\tau^{-}+\text{n-jets}\,(\text{n}\geq 2)+\slashed{E}_{T}$.  Among these, the model signature $\ell^{-}+\text{n-jets}\,(1\leq \text{n} \leq 2)$ arises due to the direct decay of LQ to a lepton and a jet.  All the other final states arise due to the decay of LQ to a RH neutrino and a light jet, or to a RH neutrino and top quark, with successive decays of RH neutrino, and $t$ quark into SM states.

We find that, among all the above mentioned final states, $\ell^{-}+\text{n-jets}\,(1\leq \text{n} \leq 2)$ has the highest discovery prospect even after giving a generic set of cuts.  A LQ of mass upto 1.4 TeV in this channel can be discovered 
at more than 5$\sigma$ C.L. with 100 $\rm{fb}^{-1}$ of data.  The LQs will also result in 
the enhancement  of the RH neutrino production in association with a light jet, or with top quark.  If at LHeC 
the 
electron beam is polarized, the right handed neutrino- light jet production 
cross-section can substantially increase~\cite{Subhadeep,Mondal:2015zba}.  We
find that among all the final states $\ell^{-} + \text{n-jets} (\text{n} \geq 3)$, and $\bar{b}\ell^+\tau^-+\text{n-jets} (\text{n}\geq 2)+\slashed{E}_T$ are the most optimal, after implementing the selection cuts judiciously. 
With $100\,\text{fb}^{-1}$ integrated luminosity, for LQ mass 1.1 TeV, the expected number of events for the final sates $\ell^{-} + \text{n-jets} (\text{n} \geq 3)$, and $\bar{b}\ell^+\tau^-+\text{n-jets} (\text{n}\geq 2)+\slashed{E}_T$ are $10^{4}$ and $180$ respectively.

{\bf{Acknowledgement}}: M.M. would like to acknowledge the DST-INSPIRE research grant IFA14-PH-99. 

\appendix

\bibliography{bibitem}

\end{document}